\newcommand{\m}{\mbox}
\documentstyle[12pt,aasms4]{article}
\newcommand{\Ms}{\m{$\,M_\odot$}}
\newcommand{\Rs}{\m{$\,R_\odot$}}
\newcommand{\Ls}{\m{$\,L_\odot$}}
\newcommand{\Msyr}{\m{$\,M_\odot\,$yr$^{-1}$}}

\newcommand{\yr}{\m{\,yr}}

\begin{document}

\title{Cygnus X-2: the Descendant of an Intermediate-Mass X-Ray Binary}
\author{Ph.\ Podsiadlowski}
\affil{Oxford University, Oxford, OX1 3RH, U.K.}
\authoremail{podsi@astro.ox.ac.uk}

\author{S. Rappaport}
\affil{Department of Physics and Center for Space Research, Massachusetts
Institute of Technology}
\authoremail{sar@eagle.mit.edu}

\begin{abstract}
The X-ray binary Cygnus X-2 (Cyg X-2) has recently been shown to
contain a secondary that is much more luminous and hotter than is
appropriate for a low-mass subgiant. We present detailed
binary-evolution calculations which demonstrate that the present
evolutionary state of Cyg X-2 can be understood if the secondary had
an initial mass of around $3.5\Ms$ and started to transfer mass near
the end of its main-sequence phase (or, somewhat less likely, just
after leaving the main sequence). Most of the mass of the secondary
must have been ejected from the system during an earlier rapid
mass-transfer phase. In the present phase, the secondary has a mass of
around $0.5\Ms$ with a non-degenerate helium core.  It is burning
hydrogen in a shell, and mass transfer is driven by the advancement of
the burning shell. Cyg X-2 therefore is related to a previously little
studied class of intermediate-mass X-ray binaries (IMXBs). We suggest
that perhaps a significant fraction of X-ray binaries presently
classified as low-mass X-ray binaries may be descendants of IMXBs and
discuss some of the implications.
\end{abstract}

\keywords{stars: binaries --- stars: evolution --- stars: individual
(Cygnus X-2) --- X-rays: stars}

\section{Introduction}
Cygnus X-2 has long been considered a typical low-mass X-ray binary
containing a neutron star \markcite{Smale}(e.g., Smale 1998)
with a relatively long orbital period
($P = 9.84\,$d; \markcite{Cowley}Cowley, Crampton \& Hutchings 1979).
Recent spectroscopic observations by \markcite{Casares}
Casares, Charles \& Kuulkers (1998) and a study of the ellipsoidal
light variation by \markcite{Orosz}Orosz \& Kuulkers (1998) yielded accurate
values for the mass ratio ($q=0.34\pm 0.04$), the individual
masses ($M_{x} = 1.78\pm 0.23\Ms$ and $M_c=0.60\pm 0.13\Ms$ for the
neutron star and the companion, respectively), and the radius
of the companion ($R_c=7.0\pm 0.5\Rs$; unpublished).
These observations confirm the {\em present} low mass of the companion. \par

Rather surprisingly, however, \markcite{Casares}Casares
et al.\ (1998) also showed that
the spectral type of the companion is, to within two subtypes, A9III
and that {\em the spectral type does not vary with orbital
phase}. Using the binary parameters above and adopting a temperature
of $\sim 7400\,$K for this spectral type \markcite{Strai}(Strai\v{z}ys
\& Kuriliene 1981), we obtain a luminosity $L_{c}\sim
130\Ls$ for the companion star. On the other hand, we can also
estimate the expected temperature and luminosity for a low-mass
subgiant that is consistent with the binary parameters of Cyg X-2:
using the calibrated stellar models by \markcite{Han}Han (1995), we
find an expected temperature of 4500$\,$K and an expected luminosity
of 18\Ls. This much higher temperature and luminosity of the companion
star could in principle be caused by X-ray heating (also see
\markcite{Cowley}Cowley et
al.\ 1979). However, the fact that the spectral type does not vary
with orbital phase despite an inclination $i\sim 62\arcdeg$
\markcite{Orosz}(Orosz \& Kuulkers 1998) would imply extremely
efficient redistribution of the irradiation energy around the star by
irradiation-induced circulation currents, much higher than is, for
example, observed in the companion of Her X-1
\markcite{Kippenhahn}(Kippenhahn \& Thomas 1979). The spectra
of \markcite{Casares2}Casares \& Charles (1998)
also show a helium absorption line whose
strength varies with orbital phase. The very
presence of this line on the companion indicates regions of much
higher temperature than found on an A9 star. Since the strength of
this line is strongest on the illuminated side of the secondary, this suggests
that it is caused by external illumination and heating, while
the non-varying component of spectral type A9 reflects the
true intrinsic spectral type of the companion.
\par
Since a spectral type A9 is not consistent with a low-mass subgiant,
this immediately suggests that the original mass of the companion must
have been much higher and that most of the mass of the companion must
have been lost from the system. Indeed, in \S 2 we will show that the
present parameters of Cyg X-2 can be understood if, at the beginning
of the mass-transfer phase, the companion was a somewhat evolved star
of mass $\sim 3.5\Ms$ that lost most of its envelope as a result of
highly non-conservative mass transfer. Cyg X-2 may therefore be
related to a hitherto little studied class of intermediate-mass X-ray
binaries (IMXBs). In \S 3 we discuss some of the
implications of this conclusion and how it may affect our
understanding of X-ray binaries in general.\par
We note that a  similar suggestion concerning the evolutionary
status of Cyg X-2 was recently made independently by King \& Ritter (1999).
However, as we will show in \S 2, our best model (case AB) differs
substantially from their suggested model (early case B), both in its details
and its predictions.

\section{Binary Calculations}

In order to explore the possibility that Cyg X-2 is the descendant of
an intermediate-mass X-ray binary, we performed a series of binary
calculations using an up-to-date, standard, Henyey-type stellar
evolution code (\markcite{Kippenhahn3}Kippenhahn, Weigert \&
Hofmeister 1967). Our calculations use solar metallicity ($Z=0.02$),
an initial hydrogen abundance of $X=0.70$, a mixing-length parameter
$\alpha=2$, and use a standard prescription for the mass-transfer rate
$\dot{M}$ (\markcite{Ritter}see, e.g., Ritter 1988). For each
binary-evolution sequence, one needs to specify what fraction,
$\beta$, of the mass lost by the donor is accreted by the neutron
star, and the specific angular momentum of any matter that is lost
from the system. We somewhat arbitrarily
choose $\beta$ to be 1/2, and limit the maximum
accretion rate to the Eddington accretion rate, taken to be $\dot{M} =
2\times 10^{-8}\Msyr$ and kept constant throughout each run.
We further assume that the mass lost from the system
carries away the specific angular momentum of the accreting neutron
star which has an initial mass of 1.4\Ms. We include angular-momentum
loss due to gravitational radiation (although it is of no real
importance here), but do {\it not} consider magnetic braking, since during
all slow evolutionary phases, the envelopes of our donor stars tend to
be fully radiative (for a general background review see, e.g.,
\markcite{Ritter2}Ritter 1996). We have also performed some calculations that
include magnetic braking and found that this would not change any of the main
conclusions in this paper.
\par

The evolution of a binary is very sensitive to the evolutionary state
of the donor star at the beginning of the mass-transfer phase. For example,
if the donor is initially relatively unevolved (early case A;
\markcite{Kippenhahn2}Kippenhahn \& Weigert 1967), it will subsequently
mimic a single star of the same actual mass rather than its initial mass
(e.g., \markcite{Hellings}Hellings 1983),
provided that is is not too far out of thermal
equilibrium. Since the companion of Cyg X-2 does not presently resemble
a single star of the same mass, this type of evolution cannot be applicable.
To illustrate this evolution, we show in Figure~1
the evolutionary track in the Hertzsprung-Russell (H-R) diagram of a 3.5\Ms\
donor star that is somewhat evolved (marked `case A'). At the beginning
of the mass-transfer phase, it has used up only half of its initial hydrogen
supply in the center. After a short, rapid mass-transfer phase (dashed
portion), the star spends most of the remaining main-sequence phase in
the low-mass region of the main sequence (dot-dashed portion)
and then evolves like a low-mass star up the Hayashi line (solid portion).
It never resembles a star like the companion
of Cyg X-2 (the parameters for Cyg X-2 are shown as a
boxed region in Fig.~1).
\par
In order to explain the presently overluminous companion, it must have
been relatively evolved at the beginning of the mass-transfer phase.
This constraint allows two possibilities: either the
secondary was near the end of the main sequence and had already developed
a core structure typical of a post-main-sequence
intermediate-mass star (case AB) or it had already left the main
sequence and filled its Roche lobe while evolving through the Hertzsprung gap
(case B).
\par
\subsection{The Case AB Scenario}
The curves marked `case AB' in Figure~1 and Figure~2 illustrate the
evolution of a 3.5\Ms\ star that fills its Roche lobe for the first
time near the end of the main sequence, when its central hydrogen
abundance has been reduced to $X_c\simeq 0.10$. The exact value of $X_c$
is not crucial, but should probably be less than $\sim 0.2$.
The panels in Figure~2 show the radius and Roche-lobe
radius (panel a), the orbital period (panel b), the masses of the two
components and the secondary's core mass (panel c),
and the mass-loss rate from the donor (panel d) since
the beginning of the mass-transfer phase. As is most evident from the last
panel, one can clearly distinguish three separate phases, one very rapid
phase and two slower phases.
\par\medskip
\noindent{\em The rapid initial phase}\par

\noindent
Because of the large initial mass ratio, the initial mass-transfer
rate is of order $10^{-5}\Msyr$, and most of the mass lost from the
companion has to be ejected from the system. The high $\dot{M}$
implies a mass-loss time scale that is short compared to the thermal
time scale of the donor, and the donor star is therefore driven
significantly out of thermal equilibrium (i.e., it is undersized and
underluminous for its mass). The mass-transfer rate only starts to
decrease significantly after the mass ratio has been reversed and the
system, and hence the Roche lobe, begin to expand. At this stage, mass
transfer is driven entirely by the thermal expansion of the companion.
The rapid phase ends
once the companion has re-established thermal equilibrium (after $\sim
2\times 10^6\,$yr, of order the thermal time scale of the star).
By this time, the secondary's mass has decreased to 0.900\Ms. Since the
material that is now exposed at the surface has undergone partial CNO burning,
its composition shows the signature of CNO processing (enhanced nitrogen,
decreased carbon and oxygen) and the surface hydrogen mass
fraction is reduced to 0.55 (from 0.7).
\par\medskip

\noindent{\em The hydrogen core-burning phase}

\noindent
Once the star has returned to thermal equilibrium, the further evolution
is driven by the nuclear evolution of the core (hydrogen core
burning). During this phase, which in this example lasts $6\times 10^7\,$yr,
the mass-transfer rate is of order $4\times 10^{-10}\Msyr$ and the mass
decreases to 0.876\Ms. The phase ends when hydrogen has been exhausted
in the core and the system becomes temporarily detached, accompanied by
a small hook in the H-R diagram (Fig.~1).
\par\medskip

\noindent{\em The hydrogen shell-burning phase}\par
\noindent
After the exhaustion of central hydrogen, the star expands again and
soon starts to fill its Roche lobe for a second time.  Initially the
evolution of the secondary resembles that of a low-mass star evolving
off the main sequence. However, its behavior changes qualitatively
once it has lost all of the material that was outside the convective
core at the beginning of the mass-transfer phase (which had a mass of
$0.5\Ms$). At this point, the surface hydrogen abundance drops to 0.1
(the central hydrogen abundance in the core at the beginning of mass
transfer), and the star now has many of the characteristics of a
non-degenerate helium star, except for the fact that the evolution,
and hence mass transfer, are driven by hydrogen burning in an
outward-moving shell.  Unlike the case of early case B (see \S~2.2),
the star has now no tendency to become a giant, which would lead to mass
transfer on a thermal time scale (we have tested this by evolving such a
star without further mass loss).  The mass-transfer rate is therefore
determined by the time scale for hydrogen-shell burning
and lies in the range of $2\times 10^{-9}\Msyr$ to $7\times
10^{-8}\Msyr$ (see the inset in Fig.~2(d)).  This phase ends when the
hydrogen-rich envelope has almost been exhausted and hydrogen-burning
stops. The star still continues to evolve, now as a detached star,
burns helium in the core as an O subdwarf (with a luminosity $\sim
10\Ls$ and temperature $\sim 30,000\,$K) and eventually ends its
evolution as a CO white dwarf with an unusually low mass of
$0.414\Ms$. The final neutron-star mass in this calculation is
$1.64\Ms$.
\par\medskip

\noindent{\em The appearance of the secondary}\par\medskip
\noindent
In the initial rapid phase (dashed portion of curve AB in Fig.~1),
most of the transferred mass has to be ejected from the system. Since
the mass-loss rate exceeds the Eddington rate by several orders of
magnitude, the system is unlikely to look like a typical X-ray
binary. Since this phase is rather short-lived ($\sim 10^6\yr$),
relatively few systems should be found in this phase. One possibly
related, somewhat more massive observed counterpart is the famous
system SS433, which appears to eject most of the transferred mass by
means of two relativistic jets (e.g., \markcite{Margon}Margon 1984).
\par
On the other hand, in the two nuclear-evolution-driven phases, the
system would in most respects resemble a typical low-mass X-ray binary
(LMXB) and would almost certainly be observationally classified as a
low-luminosity LMXB in the core-burning phase, and a high-luminosity
LMXB in the shell-burning phase. Both phases are also relatively
long-lived ($6\times 10^{7}$ and $3\times 10^{7}\yr$) and hence
relatively more systems should be found in these phases than in the
rapid phase (see \S~3 for discussion).\par
In this particular model, the star spends a large fraction of the
shell-burning phase near a location in the H-R diagram where Cyg X-2
lies (the dashed portion of track AB in Fig.~1). Indeed during the phase
when the orbital period increases from $\sim 7$ to $\sim 10\,$d, the
model provides an excellent description of the main observed properties
of Cyg X-2 (see \S~1; the mass of the secondary is in the range of 0.45
and 0.5\Ms). An important prediction of this model is that the
surface composition should be the same as the {\it core composition}
of the secondary at the beginning of the first mass-transfer phase, i.e.,
be severely hydrogen depleted (in this particular model, $X\simeq 0.1$)
and show strong evidence of CNO processing.
(We caution, however, that there are a sufficient
number of uncertainties in the binary-evolution model, in particular
associated with the angular-momentum loss from the system, that somewhat
different initial binary parameters can almost certainly also
reproduce the observed properties of Cyg X-2.)

\subsection{An Early Case B Scenario?}

An alternative model for Cyg X-2 is that it started to fill its Roche
lobe in the Hertzsprung gap (early case B; also see \markcite{Kolb}Kolb
 1998 and, in particular, \markcite{King}King \& Ritter 1999).
We have calculated this type of evolution for a 3.5\Ms\ star
that has just left the main sequence (dashed curve marked `case B' in
Fig.~1). The initial evolution is similar to the previous case.  The
mass-transfer rate (dashed curve in Fig.~2(d)) reaches a peak of $\sim
10^{-5}\Msyr$ and starts to drop once the mass ratio has been
reversed. In the subsequent somewhat, but not much slower phase,
mass-transfer is driven by
the thermal evolution of the envelope (the star is trying to evolve
across the Hertzsprung gap to become a giant). This phase ends and
mass transfer stops completely once most of the hydrogen-envelope has
been lost and the hydrogen-burning shell has been extinguished. The secondary
continues to evolve as a detached star, burning helium as an O
subdwarf and finally becomes a CO white dwarf of mass $0.617\Ms$ (the
neutron star has only a slightly increased final mass of $1.427\Ms$).

As is clear from Figure~2(d), apart from the initial phase, the
mass-transfer history is very different from the case AB scenario
above. In the slower phase, mass transfer is driven by the thermal
expansion of the secondary rather than the nuclear evolution of the
core or shell.  Since the thermal time scale is much shorter than the
evolutionary time scale during the hydrogen-shell burning phase in the
case AB scenario, the mass-transfer rate in the case B scenario
is $\sim 3\times 10^{-7}\Msyr$, an order of magnitude above the Eddington
rate.  It is not clear
whether such a high mass-transfer rate would be consistent with the
observed properties of Cyg X-2. This problem is somewhat reduced if
the initial secondary mass was somewhat lower ($\sim 3\Ms$).  However,
all case B scenarios suffer from the generic problems that the
mass-transfer rate is significantly super-Eddington and that the
characteristic lifetime of this phase is very short (of order the
thermal time scale of the star, $\sim 10^6\yr$). In all the case B
models we have
calculated, the secondaries spend very little time in the H-R diagram
in the general neighborhood of Cyg X-2. For even lower masses, the
thermal time scale would be longer, but the star would have a tendency
to evolve towards the Hayashi line, which again would be inconsistent with
the inferred parameters of the secondary in Cyg X-2.  We have performed a
whole series of early-case-B binary sequences.  None of them was
able to reproduce all of the observationally inferred parameters of
Cyg X-2 simultaneously (i.e., the position in the H-R diagram,
the constituent masses, and $\dot{M}$).
While we cannot strongly rule out an early-case-B scenario
for Cyg X-2, keeping in mind the general uncertainties in modeling
non-conservative mass transfer, we conclude that it is more likely
that Cyg X-2 is at present in a hydrogen-shell burning phase in a case
AB scenario.

\section{The Importance of Intermediate-Mass X-Ray Binaries (IMXBs)}

An important implication of our modeling of Cyg X-2 is that it
suggests that there may be a whole class of X-ray binaries which
either have intermediate-mass secondaries now or had them in the past.
Such systems have received hardly any attention in the past (see, however,
\markcite{Pyl1}Pylyser \& Savonije 1988, \markcite{Pyl2}1989;
\markcite{Kolb}Kolb 1998). To some degree, this is the result of
observational findings, since previously only one such system (Her
X-1/HZ Her) had been unambiguously identified.  In addition, the
theoretically predicted large initial mass-transfer rates are much
higher than those typically inferred for observed X-ray binaries.
However, as our calculation show, this rapid phase is relatively
short-lived and very few systems should be found in this phase. It is
also not clear what they would look like. A further theoretical
uncertainty is how the mass is lost in this rapid phase. As noted
earlier, SS433 may provide an observational answer to both of these
questions (the appearance initially and the mass-loss mechanism). On
the other hand, from an evolutionary point of view, IMXBs should be
relatively common, since they are much easier to form and do not
require the same of amount of fine-tuning as LMXBs (see, for example,
\markcite{Bhatta}Bhattacharya \& van den Heuvel 1991). The
observational properties of IMXBs would vary significantly, depending
on the initial mass of the secondary and the evolutionary phase at the
beginning of the mass-transfer phase.

We have initiated a systematic
investigation of IMXBs and found that, in the longer-lasting
evolutionary phases, they typically resemble `classical' LMXBs
(\markcite{Podsi1}Podsiadlowski
\& Rappaport 1999). This suggests that perhaps a significant
fraction of X-ray binaries, presently classified as LMXBs, are
actually IMXBs or their descendants. This may have important
implications for our understanding of X-ray binaries. For example, it
may help to explain some of the systematic differences between
Galactic and globular-cluster LMXBs, the observed period and
luminosity distributions, and may help to resolve the apparent
discrepancy between the number of millisecond pulsars and their
putative progenitors (e.g., \markcite{Kulkarni}Kulkarni \& Narayan 1988).

\section{IMXBS and millisecond pulsars}

        There is a group of about two dozen radio pulsars that are
found in nearly circular orbits with low-mass white-dwarf companions
(see, e.g., \markcite{Taylor}Taylor, Manchester \&
Lyne 1993; \markcite{Rappaport}Rappaport et al.\ 1995).
 The orbital periods of these systems range from less than 1 day to
1232 days.  It has been conjectured that this type of system evolves
from a low-mass ($\sim 1\Ms$) subgiant or giant donor star which
transfers its envelope to a neutron star via stable Roche-lobe
overflow (see, e.g., \markcite{Joss}Joss, Rappaport \& Lewis 1987;
\markcite{Rappaport}Rappaport et al.\ 1995).  What remains is a
low-mass white dwarf (the core of the
giant) in a wide, nearly circular orbit about the neutron star which
has been spun up by the accretion process.

\markcite{Joss}Joss, Rappaport, \& Lewis (1987) and
\markcite{Rappaport}Rappaport et al.\ (1995) showed
that there is a nearly unique relation between the final orbital period,
$P$, of such a ``recycled'' binary pulsar and the mass of the white dwarf,
$M_{\rm wd}$ (see also \markcite{Savonije}Savonije 1987, and
\markcite{Refsdal1}Refsdal \& Weigert 1970, \markcite{Refsdal2}1971 for
related discussions).  In short, this relation results from the fact that
the radius of a low-mass giant is a nearly unique function of its core
mass.  It then follows for a Roche-lobe filling star in a binary that $P$,
at the end of mass transfer, is a function of only the core mass of the
donor.  Thus, the binary remains as a fossil relic of the giant donor star
and its degenerate He core.  The theoretical relation between $P$ and
$M_{\rm wd}$ is found to be roughly consistent with the inferred masses of the
white dwarf companions in these systems.  Unfortunately, this model cannot
be confirmed to the level of confidence that one would like.  First, for
most of these systems, only the mass function is measured, and thus the
uncertainty in the white dwarf mass is rather large.  Second, there are a
number of these systems for which the orbital period is really too short to
be explained by this scenario (i.e., $< 3\,$d).  Finally, there are a few
systems where the lower limit on the mass of the white dwarf is only
marginally consistent with the theoretical relationship.

        The last of these difficulties could be mitigated if some of these
systems formed with donors of intermediate mass.  We have shown in this
work that neutron stars with intermediate-mass donor stars can attain a
final evolutionary state which is very much like that of the binary radio
pulsars discussed here.  In particular, the system which we use to model
Cyg X-2 would end its evolution with a 0.42\Ms\ CO white dwarf in a nearly
circular 12-day orbit about a spun-up neutron star.  Since the initial mass
transfer in this system occurred when the 3.5\Ms\ donor star was late in
its main-sequence phase (late case A), the core was neither completely
hydrogen depleted nor degenerate.   This combination enables the final
orbital period to remain relatively short for a white-dwarf mass as large
as 0.42\Ms.  By contrast, in the more standard low-mass giant scenario the
corresponding final orbital period with this same white dwarf mass would
have been $\sim 400$ days.  Put another way, in this same scenario, a
system with a 12-day orbital period would be expected to have $M_{\rm wd}\sim
0.21\Ms$.  In the future, deep spectroscopy of the companion white dwarfs
(see, e.g., \markcite{Kerk}van Kerkwijk \& Kulkarni 1997), as well as the
possible
detection of the Shapiro delay in the arrival times of the radio pulses
(see, e.g., \markcite{Kaspi}Kaspi, Taylor, \& Ryba 1994), could
substantially improve the
determination of the white dwarf masses and thereby discriminate between
low-mass and intermediate-mass progenitors.

        It is also possible that neutron stars accreting from intermediate
mass donor stars could attain a final evolutionary state consisting of a
very low-mass white dwarf in a short orbital period binary (e.g., $< 1\,$d),
and help explain binary radio pulsars with such properties.  We plan to
explore such scenarios in future work.

        Thus, evolutionary scenarios and population synthesis studies which
are employed to reproduce the distribution of orbital periods and white
dwarf masses in binary radio pulsars, must include consideration of
intermediate mass donor stars.

\section{Conclusions}
We have shown that the  observationally inferred parameters of the
X-ray binary Cyg X-2 can be understood if the secondary had an initial
mass of $\sim 3.5\Ms$ and started to transfer mass near the end
of the main sequence (or less likely just after leaving the main
sequence). Cyg X-2 is therefore related to a class of intermediate-mass
X-ray binaries that has been little studied before. Our
favored model for Cyg X-2 (case AB) predicts that the secondary should,
at present, be severely hydrogen depleted. We suggest that perhaps
a significant fraction of X-ray binaries, that are presently classified
as LMXBs, could be related to this class.  This could have far-reaching
implications for our understanding of X-ray binaries. Detailed future
observations of the secondaries in X-ray binaries as well as systematic
binary population synthesis studies should help to assess the importance
of this relatively unexplored evolutionary channel.

\acknowledgments {The authors are grateful to P. Charles and D. Chakrabarty
for stimulating discussions.  This research was supported in part by
NASA ATP Grant NAG5-4057.}

\eject
\centerline{\bf Figure Captions}
\par\bigskip\bigskip
\noindent{\bf Figure 1.} Evolutionary tracks of the secondaries in
three binary calculations in the Hertzsprung-Russell diagram.  The
secondary has an initial mass of $3.5\Ms$ and the primary, assumed to
be a neutron star, an initial mass of $1.4\Ms$ in all calculations. The
dotted curve shows the track of a $3.5\Ms$ star without mass
loss. The mass-loss tracks, labelled case A, AB and B, start at
different evolutionary phases of  the secondary (`case A': the middle
of the main sequence; `case AB': the end of the main sequence;
`case B': just after the main-sequence). The dashed portions in each track
indicate the rapid initial mass-transfer phase, the dot-dashed and
solid portions  the slow phases where mass-transfer is driven by hydrogen
core burning and hydrogen shell burning, respectively (only in case A
and case AB). The beginning and end points of the various phases are
marked by solid bullets, the small figures next to them give the
mass of the secondary at these points. The boxed region labelled `Cyg'
indicates the observationally determined parameter region for the secondary
in Cyg X-2. The tracks after mass transfer has ceased are not shown.

\par\bigskip\bigskip
\noindent{\bf Figure 2.} Key binary parameters for the case AB binary
calculation as a function of time (with arbitrary offset). Panel (a):
radius (solid curve) and Roche-lobe radius (dot-dashed curve) of the
secondary; panel (b): the orbital period (solid curve); panel (c):
the mass of the secondary (solid curve), of its hydrogen-exhausted
core (dotted curve), and of the primary (dot-dashed
curve); panel (d): the mass-loss rate from the secondary (solid
curve); the inset shows a blow-up of the second slow mass-transfer phase
(hydrogen shell burning). The dashed curves in panels (b) and (d)
show the orbital period and mass-transfer rate for the case B calculation
for comparison.

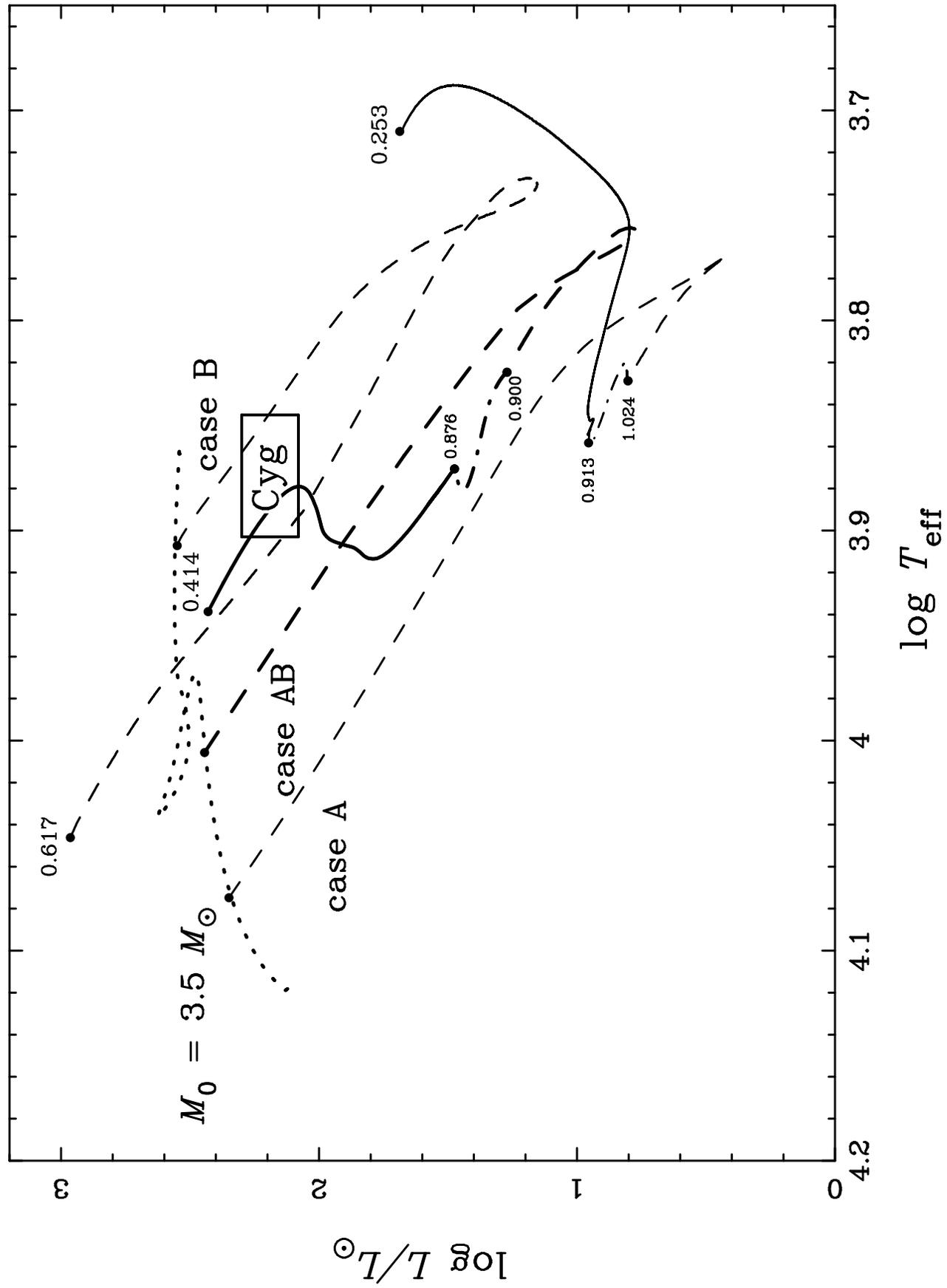
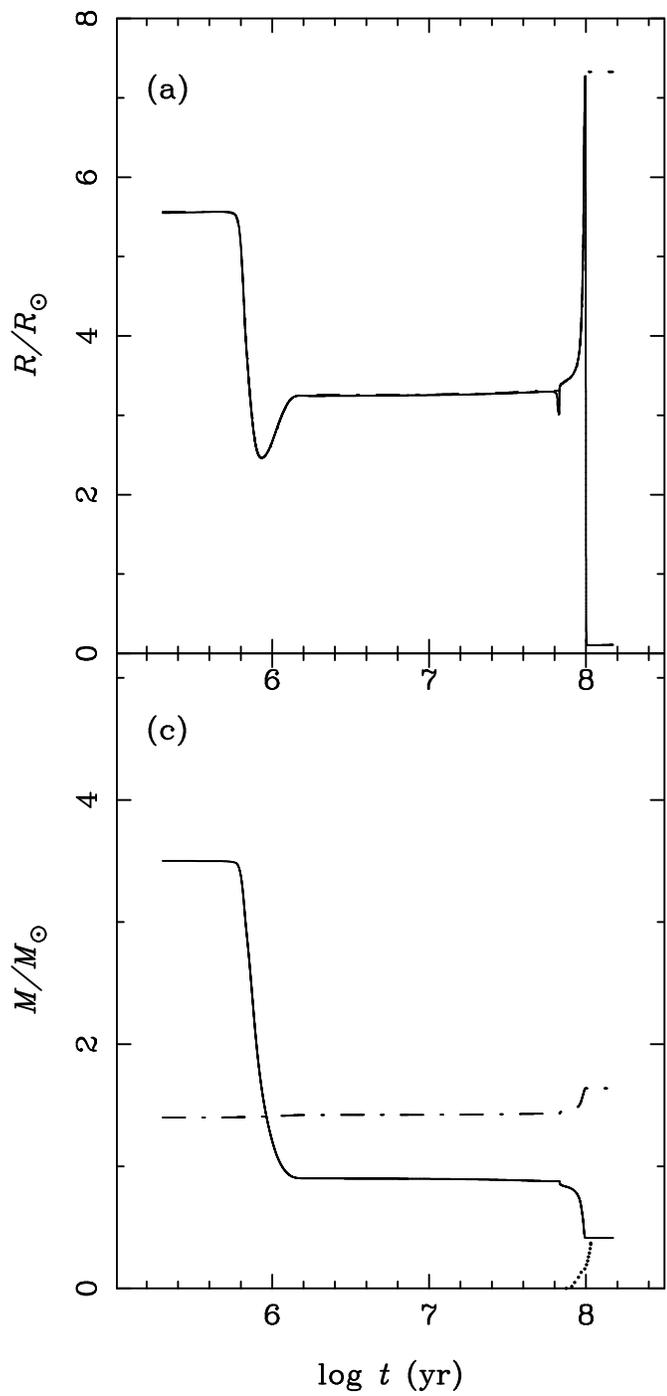
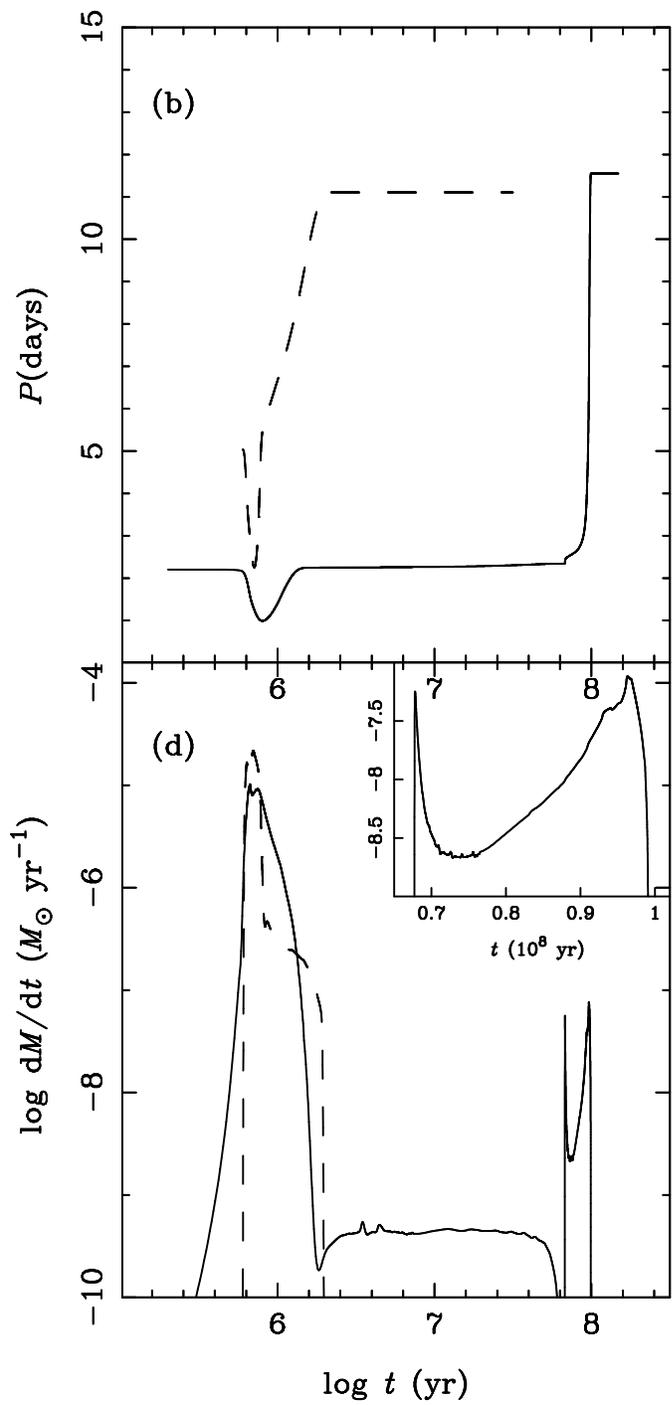
\end{document}